\documentclass[runningheads]{llncs}
\usepackage[T1]{fontenc}
\usepackage{graphicx}
\usepackage{booktabs}
\usepackage[misc]{ifsym}
\newcommand{\corr}{(\Letter)}
\usepackage{amsmath} 
\usepackage{multirow}
\usepackage{amsfonts}
\usepackage{mwe}

\begin{document}

\title{RAE: A Rule-Driven Approach for Attribute Embedding in Property Graph Recommendation}
\tocauthor{Sibo Zhao, Michael Bewong, Selasi Kwashie, Junwei Hu, Zaiwen Feng}
\toctitle{RAE: A Rule-Driven Approach for Attribute Embedding in Property Graph Recommendation}

\titlerunning{A Rule-Driven Approach for Attribute Embedding}

\author{Sibo Zhao\inst{1} \and  
Michael Bewong\inst{2,3} \and
Selasi Kwashie\inst{2} \and
Junwei Hu\inst{1} \and
Zaiwen Feng\inst{1,4,5,6}\corr}

\authorrunning{S. Zhao et al.}

\institute{
College of Informatics, Huazhong Agricultural University, Wuhan, China \\
\email{Zaiwen.Feng@mail.hzau.edu.cn}
\and
AI \& Cyber Futures Institute, Charles Sturt University, Bathurst, Australia 
\and
School of Computing, Mathematics \& Engineering, Charles Sturt University, Wagga Wagga, Australia 
\and
Hubei Key Laboratory of Agricultural Bioinformatics, Wuhan, China
\and
Engineering Research Center of Agricultural Intelligent Technology, Ministry of Education, Wuhan, China
\and
Hubei Three Gorges Laboratory, Wuhan, China
}

\maketitle              

\begin{abstract}
Recommendation systems are crucial in modern applications to enhance the user experience and drive business conversion rates through personalization. However, insufficient utilization of attribute information within the property graph remains a significant challenge. Most existing graph convolutional network (GCN) models do not consider attribute information, and those that do often employ a simplified triple format <users, items, attributes>, which fails to fully exploit the rich semantic structures of property graphs necessary for effective recommendations. To overcome these limitations, we introduce Rule-Driven Approach for Attribute Embedding (RAE), a novel methodology that enhances recommendation performance by effectively mining and utilizing semantic rules from property graphs. RAE applies a rule-mining process to extract meaningful rules that guide random walks in generating enriched attribute embeddings. These enriched embeddings are subsequently integrated into GCNs, surpassing conventional triple-based embedding techniques. We evaluate RAE on real-world datasets (e.g., Blogcatalog and Flickr) and demonstrate that RAE achieves an average improvement of 10.6\% in both Recall@20 and NDCG@20 compared to state-of-the-art baselines, indicating superior relevance coverage and ranking rationality in top-20 recommendations. Additionally, RAE exhibits enhanced robustness against data sparsity and the attribute missingness problem. Our novel approach underscores the significant performance gains achieved in recommendation systems by fully leveraging attribute information within property graphs, enhancing both effectiveness and reliability.

\keywords{Recommendation Systems  \and Property Graphs \and Rule Mining \and Attribute Embedding \and Graph Convolutional Network}
\end{abstract}

\section{Introduction}
Recommendation systems play an essential role in modern online applications, enhancing user experience by providing personalized content and services. These systems are crucial in e-commerce, social media, and entertainment platforms, helping users discover relevant items and mitigate information overload.
Traditional recommendation algorithms primarily depend on users' historical behaviors, such as clicks and purchases, to predict preferences. However, as the volume and diversity of data continue to grow, conventional approaches increasingly encounter challenges in providing accurate and relevant recommendations.

In recent years, Graph Convolutional Networks (GCNs) have gained popularity due to their capacity to model complex relationships within user-item interaction networks, which have shown promising results in recommendation systems. Nevertheless, existing GCN-based models still suffer from the data sparsity problem. When there are limited user reviews or insufficient social information, such data sparsity restricts the models' ability to capture detailed user preferences, thereby reducing the effectiveness of recommendations.

\begin{figure}[t]
\centering
\includegraphics[width=\textwidth]{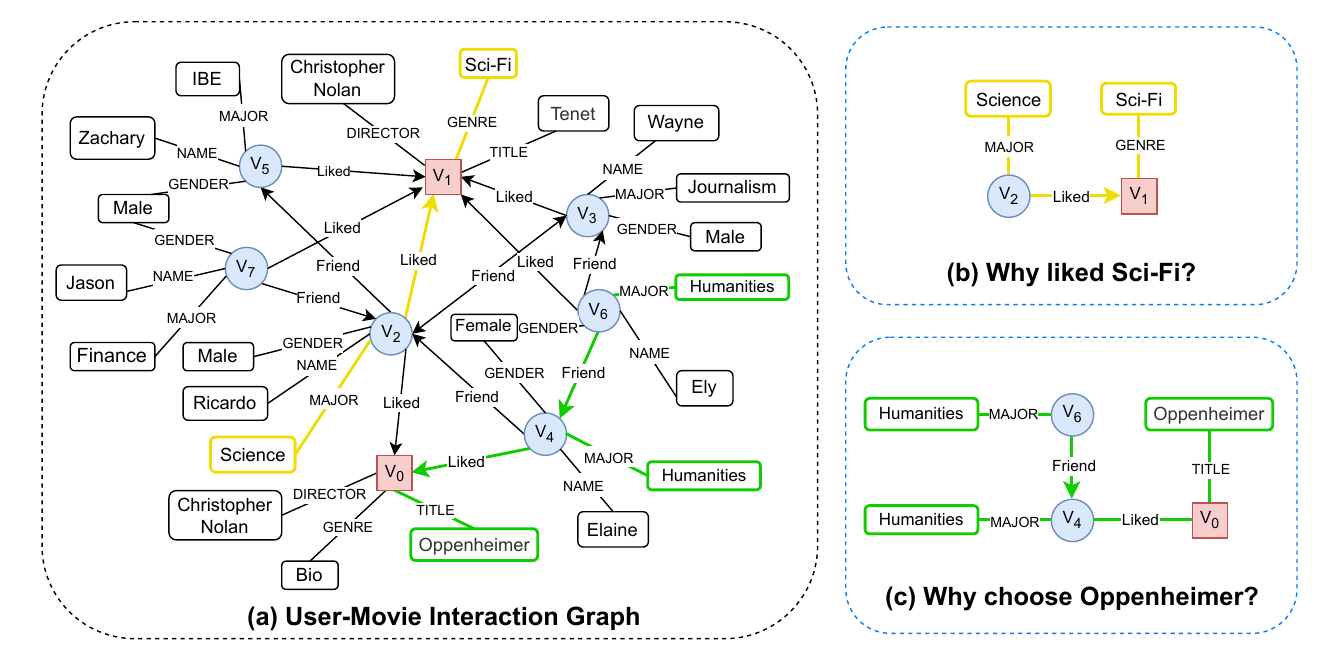}
\caption{Illustration of motivation of the proposed RAE} \label{fig:graph}
\end{figure}

To address data sparsity, researchers have explored the use of attributes as valuable side information. Attributes provide critical supplementary information in recommendation systems, especially for understanding user preferences and item characteristics, offering additional insights that can improve recommendation accuracy. Despite their potential, current attribute-enhanced models often underutilize these attributes \cite{a2gcn,afgcn}, treating them merely as simple triples of the form \texttt{<users, items, attributes>}. Such representations do not fully exploit the rich semantic structures of attributes embedded within property graphs, where attributes originate from and can convey complex relationships and semantic meanings. 

Motivated by the limitations of existing recommendation models in handling sparse and incomplete attribute data, we propose a novel \textbf{Rule-Driven Attribute Embedding (RAE)}\footnote{For more details: \url{https://github.com/Sibo-Zhao/RAE}} framework. RAE seeks to enhance recommendation performance by effectively mining and utilizing semantic rules from property graphs. In our work, a rule represents a dependency relationship between attributes in the graph, such as \( x.a \rightarrow y.b \), indicating that attribute \( a \) of node \( x \) determines attribute \( b \) of node \( y \). For example, Fig. \ref{fig:graph}(a) represents a user-movie interaction graph, from which 
we can derive the rule \( V_2.\text{MAJOR} \rightarrow V_1.\text{GENRE} \) (shown in Fig. \ref{fig:graph}(b)), implying that user \( V_2 \)'s major in Science influences his preference for the Sci-Fi genre represented by movie \( V_1 \). Similarly, the rule \( (V_6.\text{MAJOR} = V_4.\text{MAJOR}\) \( \land \) \(V_4.\text{MAJOR} = \text{Humanities} \rightarrow V_0.\text{TITLE} )\) (Fig. \ref{fig:graph}(c)) suggests that the shared major in Humanities between users \( V_6 \) and \( V_4 \), along with their friendship, supports recommending the movie `Oppenheimer' to user \( V_6 \). 
These rules capture the semantic dependencies between user attributes and movie attributes, which can enable a recommendation system to leverage attribute information meaningfully and ultimately produce more personalized and explainable recommendations.

Our approach begins with a rule mining process to extract relevant rules, which guide random walks for generating enriched attribute embeddings. These embeddings capture deeper semantic structures than conventional triple-based representations and are integrated into GCNs to improve recommendation quality. Our main contributions are as follows:

\begin{itemize}
    \item We propose the Rule-Driven Attribute Embedding (RAE) framework, which encodes attribute information within property graphs to leverage rich semantic structures.
    \item We design a rule-based random walk process that generates attribute embeddings with enhanced semantic depth, outperforming traditional triple-based techniques.
    \item We integrate enriched attribute embeddings into GCNs, and conduct comprehensive evaluations on four real-world datasets, comparing RAE against five contemporary baselines. The results validate the framework’s effectiveness, particularly in handling challenges of data sparsity and attribute missingness.
\end{itemize}

\section{Related Work}
\subsection{Rule Discovery}
In graph-structured data, rule discovery has often concentrated on pattern-based rules related to nodes, 
similar to graph pattern mining. Examples include graph evolution rules~\cite{ger1,ger2}, link formation rules \cite{lfr}, and predictive graph rules \cite{pgr}, 
which are commonly mined using techniques such as gSpan \cite{gspan}. Advanced models like RNNLogic \cite{rnnl} focus on learning path-based rules but lack logical conditions related to attributes, limiting their ability to capture complex dependencies.
In our approach, we aim to leverage attribute-related logical conditions to enhance rule discovery in graph-based recommendation systems.
This approach is similar to methods like graph association rules (GARs) \cite{gar1,gar2}, graph functional dependencies (GFDs) \cite{gfd}, graph differential dependencies (GDDs) \cite{gdd,gddgnn}, and graph entity dependencies (GEDs) \cite{fastaged,ged2}, which integrate both graph patterns and logic conditions through either mining-based levelwise traversal \cite{gfd,gdd} or learning-based rule generation \cite{cdc}. 
The integration of logic conditions allows for more expressive rules that can capture complex relationships and dependencies, which are especially beneficial in recommendation scenarios.

\subsection{Property Graph Embedding}
Existing property graph embedding methods primarily rely on factorization- and auto-encoder-based techniques. 
Factorization-based methods like BANE \cite{bane} and LQANR \cite{lqanr} improve storage efficiency by learning binary or low-bit-width embeddings, at the cost of accuracy. 
In contrast, auto-encoder-based methods use neural networks to learn embeddings by minimizing reconstruction loss, with various network structures and proximity matrices. Recent innovations, including SAGES~\cite{sages} and PANE \cite{pane1,pane2}, aim to capture higher-order proximities and incorporate attribute data in property graphs.
Of particular interest is PANE, which is similar to our approach in treating attributes as additional nodes within the graph for random walk-based embedding generation. 
Our approach extends this concept by incorporating logic rules during the random walk process, allowing the generation of more semantically enriched and contextually informed embeddings. 

\subsection{Graph-based Recommendation}

Recent advancements in contrastive learning have significantly improved recommender systems under sparse interaction settings. Methods such as LightGCL~\cite{lightgcl}, SSLRec~\cite{sslrec}, and CGCL~\cite{cgcl} apply data augmentation and self-supervised objectives over user-item bipartite graphs. However, they typically ignore the semantic structure from attribute information. In contrast, our RAE framework operates on property graphs, extracting semantic rules to guide attribute embedding, which is orthogonal to contrastive paradigms and potentially complementary.

Knowledge graph-based models, including RippleNet~\cite{ripplenet}, KGCN~\cite{kgcn}, and RuleRec~\cite{rulerec}, enrich recommendation by leveraging external KGs via multi-hop propagation or symbolic rule induction. These methods depend on external resources and often require domain-specific construction. In contrast, RAE exploits attribute dependencies from the dataset’s internal property graph, avoiding reliance on external KGs while enabling interpretable rule-based embedding.

Graph Convolutional Networks (GCNs) have proven highly effective in learning representations from non-Euclidean structures \cite{ngcf,gcn}, making them increasingly popular in recommendation systems. LightGCN \cite{lightgcn} streamlines GCN operations by omitting feature transformation and non-linear activations, optimizing for recommendation tasks. IMP-GCN \cite{impgcn} utilizes high-order graph convolutions within subgraphs to filter out irrelevant neighbor effects, enhancing embedding relevance. Attribute-aware GCNs such as AF-GCN~\cite{afgcn} and A\(^2\)-GCN~\cite{a2gcn} integrate attributes through attention or unified graphs. Our work complements these models by incorporating rule-mined attribute embeddings into GCNs, addressing both data sparsity and missing attributes through structured semantic enrichment.

\section{Preliminaries}

\subsection{Basic Definitions and Notions}
\subsubsection{Property Graph:}
Let \( G = (V, E_v, R, E_R) \) denote a property graph, where \( V \) is the set of nodes with cardinality \( n \), \( E_v \) is the set of edges with cardinality \( m \) linking pairs of nodes in \( V \), \( R \) represents the set of attributes with cardinality \( d \), and \( E_R \) is the set of node-attribute associations. Each element in \( E_R \) is a tuple \( (v_i, r_j, w_{i,j}) \), where \( v_i \in V \) (e.g., \( V_2 \) in Fig. \ref{fig:graph}(a)) is directly associated with attribute \( r_j \in R \) (e.g., ``Science'' in Fig. \ref{fig:graph}(a)) and \( w_{i,j} \) (e.g., ``MAJOR'' in Fig. \ref{fig:graph}(a)) denotes the weight of this association, representing the attribute value. For generality, we assume \( G \) is a directed graph; if \( G \) is undirected, each undirected edge \( (v_i, v_j) \) in \( G \) is treated as two directed edges, \( (v_i, v_j) \) and \( (v_j, v_i) \), in opposite directions.

\subsubsection{Graph Patterns:}
\begin{figure}[h]
  \centering
  \includegraphics[width=\textwidth]{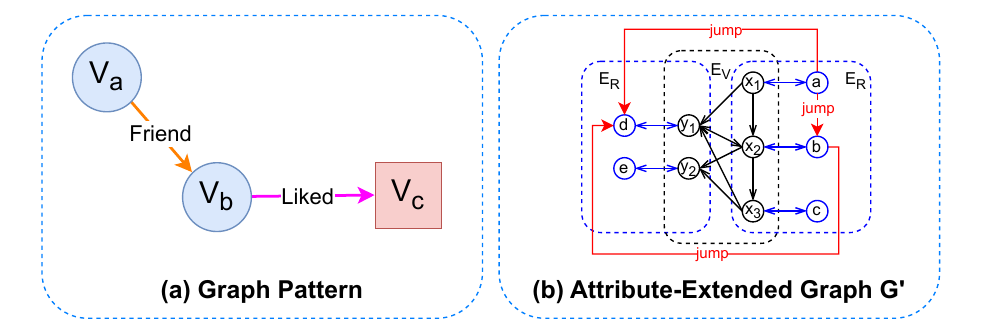}
  \caption{A Graph Pattern and Attribute-Extended Graph \( G' \)}
  \label{fig:rw}
\end{figure}

A graph pattern is defined as \( Q[\bar{x}] = (V_Q, E_Q, L_Q, \mu) \), where: (1) \( V_Q \) and \( E_Q \) denote a set of pattern nodes and edges, respectively; (2) \( L_Q \) is a function assigning each node \( u \in V_Q \) a \textbf{label} \( L_Q(u) \in L \), where \( L \) is the set of possible labels, which are distinct from attributes. (3) \( \bar{x} \) is an ordered list of distinct variables; and (4) \( \mu \) is a bijective mapping from \( \bar{x} \) to \( V_Q \), meaning that each variable in \( \bar{x} \) uniquely corresponds to a node \( v \in V_Q \). The wildcard symbol ‘\(\_\)’ is permitted as a special label in \( Q[\bar{x}] \). 
For example, consider a pattern in Fig. \ref{fig:rw}(a) where a user $ v_a $ is connected via a \text{friend} relationship to another user $ v_b $, who in turn \text{likes} an entity $ v_c $. 

\subsubsection{Literals:}
A literal in a graph pattern \( Q[\bar{x}] \) can take one of the following forms:
\begin{itemize}
    \item \textbf{Attribute literal} \( x.A \): Specifies that the variable \( x \) has an attribute \( A \).
    \item \textbf{Edge literal} \( \iota(x, y) \): Denotes an edge labeled \( \iota \) directed from \( x \) to \( y \).
    \item \textbf{ML literal} \( \mathcal{M}(x, y, \iota) \): Represents the prediction of an ML classifier regarding the existence of an edge labeled \( \iota \) from \( x \) to \( y \).
    \item \textbf{Variable literal} \( x.A = y.B \): Indicates that attribute \( A \) of \( x \) is equal to attribute \( B \) of \( y \).
    \item \textbf{Constant literal} \( x.A = c \): Asserts that attribute \( A \) of \( x \) is equal to a constant value \( c \).
\end{itemize}

\subsubsection{Rules in Graph:}
A rule in a graph \( \varphi \) is defined as: $Q[\bar{x}](X \rightarrow Y)$,
where \( Q[\bar{x}] \) denotes a graph pattern, and \( X \) and \( Y \) represent conjunctions of literals within \( Q[\bar{x}] \). We refer to \( Q[\bar{x}] \) as the \textit{pattern} and \( X \rightarrow Y \) as the \textit{dependency} of \( \varphi \). For instance, consider a rule (Fig. \ref{fig:graph}(c)) under Graph Pattern Fig. \ref{fig:rw}(a) \((V_a.major = Humanities \land V_a.major = V_b.major \implies V_c.Title = Oppenheimer)\) , suggesting that the common major in Humanities shared between users \( V_6 \) and \( V_4 \), alongside their established friendship, provides justification for recommending the film ``Oppenheimer'' to user \( V_6 \).

\subsection{Problem Formulation}

In this work, our goal is to effectively leverage the rich attribute information within property graphs to improve recommendation systems. The input consists of a set of user-item interactions \( O^+ = \{(u, i) \mid u \in U, i \in I\} \) where \( U = \{u_1, \dots, u_M\} \) and \( I = \{i_1, \dots, i_N\} \) represent the sets of users and items, respectively, and a property graph \( G = (V, E_v, R, E_R) \). Here, we focus on implicit feedback from interaction data, constructing a user-item interaction graph \( G^I = \{(u, r_{ui}, i) \mid (u, i) \in O^+\} \), where \( r_{ui} \) represents the interaction relation between user \( u \) and item \( i \).
For instance, consider a scenario where the goal is to predict whether a user will like a movie. Using IMDB data containing user interactions with movies in the form of likes, we can generate a user-item interaction graph similar to the one shown in Fig. \ref{fig:graph} (a). We can also rely on publicly available knowledge bases like Wikipedia to generate a property graph that represents movies under consideration. While existing work only relies on the information within the user-item interactions, our goal is to learn relevant rules from the property graph (\emph{i.e.}, Wikipedia) to enhance the learning of a recommendation model from the user-item interactions (\emph{i.e.}, IMDB).

In the property graph \( G \), the node set \( V \) includes both users and items, along with their associated attribute information. By integrating the user-item interaction graph \( G^I \) with the property graph \( G \), we aim to effectively utilize attributes to improve recommendation outcomes.

\begin{definition}[Problem Definition]
Given a property graph \( G \), our framework aims to develop an improved recommendation system that leverages rules to recommend items \( I'_u \) to each user \( u \) based on both the enriched embeddings and the interaction history \( I_u \). 
\end{definition}

\section{Methodology}
In this section, we introduce our novel recommendation framework, \textbf{Rule-Driven Approach for Attribute Embedding (RAE)}, designed to fully leverage attribute information in property graphs. 
Our framework is composed of three main modules: (1) a rule mining module, which extracts significant attribute association rules to identify key relationships within the graph; (2) a rule-based random walk embedding module, where the mined rules guide random walks on the property graph to generate enriched node and attribute embeddings; and (3) a recommendation module, which integrates the generated embeddings into a GCN to improve recommendation accuracy. In the following, we detail the design and implementation of each module.

\subsection{Rule Mining Module}

Our approach leverages Graph Association Rules (GAR) \cite{gar2} to extract relevant attribute association rules within property graphs. Specifically, we focus on three categories of literals: Attribute literals (\( x.A \)), Variable literals (\( x.A = y.B \)), and Constant literals (\( x.A = c \)), which encapsulate essential attribute-based relationships critical for generating meaningful embeddings. By incorporating these literal types, we emphasize the effective application of these rules to create structured associations, rather than detailing the full inference mechanisms of GAR. These mined rules subsequently guide random walks, enriching attribute embeddings and enhancing recommendation performance.

\subsection{Rule-Based Random Walk Embedding}
To fully utilize the rich attribute information within property graphs, we introduce a rule-based random walk embedding approach. Our approach builds upon the concept of random walks in attributed network embedding \cite{pane1}, further guiding these walks using mined rules to capture deeper semantic relationships within the property graph. 

\subsubsection{Attribute Matrix:} \label{sec:AM}
Given a limited embedding dimension \( k \ll n \), each node \( v \in V \) is represented by a vector of dimension \( k \) to establish a \textbf{node embedding}. The primary aim of our property graph embedding approach is to construct an embedding \( X_v \) for each node \( v \), effectively capturing both the structural connections and the attributes associated with \( v \). In line with prior research \cite{pane1}, we designate a portion of the embedding space, specifically \( \frac{k}{2} \), for each attribute \( r \in R \). This allocation enables the formation of an \textbf{attribute embedding vector} of length \( \frac{k}{2} \) per attribute, allowing our framework to efficiently integrate node interactions and attribute relations within the constrained embedding space.

In our work, matrices are represented by bold uppercase letters, such as \(\mathbf{M}\). Specifically, \(\mathbf{M}[v_i]\) indicates the row vector of \(\mathbf{M}\) corresponding to \( v_i \), while \(\mathbf{M}[:, r_j]\) refers to the column vector associated with \( r_j \). The entry at the intersection of the \( v_i \)-th row and \( r_j \)-th column in matrix \(\mathbf{M}\) is denoted as \(\mathbf{M}[v_i, r_j]\). Additionally, for any index set \( S \), \(\mathbf{M}[S]\) designates the submatrix that includes rows or columns as specified by \( S \).

Let \(\mathbf{A}\) denote the adjacency matrix for the property graph \( G \), where \(\mathbf{A}[v_i, v_j] = 1\) if there exists an edge \( (v_i, v_j) \in E_v \), and \(\mathbf{A}[v_i, v_j] = 0\) otherwise. We define \(\mathbf{D}\) as the diagonal matrix of out-degrees for \( G \), where each diagonal entry \(\mathbf{D}[v_i, v_i] = \sum_{v_j \in V} \mathbf{A}[v_i, v_j]\) represents the out-degree of node \( v_i \). The random walk matrix \(\mathbf{P} = \mathbf{D}^{-1} \mathbf{A}\) is then constructed to represent the transition probabilities for a random walk moving from node \( v_i \) to \( v_j \).

To capture attribute-level interactions, we introduce an \textbf{attribute matrix} \(\mathbf{R} \in \mathbb{R}^{n \times d}\), where each entry \(\mathbf{R}[v_i, r_j] = w_{i,j}\) encodes the weight of association between node \( v_i \) and attribute \( r_j \), as specified in the property graph edge \( (v_i, r_j, w_{i,j}) \in E_R \). The vector \(\mathbf{R}[v_i]\) is termed the \textbf{attribute vector} of node \( v_i \). Using \(\mathbf{R}\), we derive row-normalized and column-normalized attribute matrices, denoted by \(\mathbf{R}_r\) and \(\mathbf{R}_c\), respectively, defined as follows:

\begin{equation} \label{eq:nor}
\mathbf{R}_r[v_i, r_j] = \frac{\mathbf{R}[v_i, r_j]}{\sum_{r_l \in R} \mathbf{R}[v_i, r_l]}, \quad
\mathbf{R}_c[v_i, r_j] = \frac{\mathbf{R}[v_i, r_j]}{\sum_{v_l \in V} \mathbf{R}[v_l, r_j]}.
\end{equation}

\subsubsection{Rule-based Random Walk:}
Our approach leverages an \textit{extended graph} \( G' \) ( Figure~\ref{fig:rw}(b)), which is an augmented version of the original graph \( G \). The extended graph \( G' \) is constructed by adding supplementary nodes and edges to \( G \), which represent attribute associations and relationships. For example, Figure~\ref{fig:rw}(b) illustrates the extended graph \( G' \) constructed from a given property input, where the black part represents the original graph \( G \) and the blue part shows the added attribute associations \( E_R \) within \( G' \).
The embedding generated for each vertex \( v \in V \) encodes its \textit{affinity} to the attribute set \( R \), considering both attributes directly attached to \( v \) and those reachable through longer paths in \( E_V \). To capture such multi-hop relationships, we employ a random-walk-with-restart (RWR)~\cite{rw1,rw2} method specifically adapted to our directed extended graph \( G' \). Because edge orientations matter, the procedure yields two complementary affinity scores: a forward measure, denoted \( \mathbb{F} \), and a backward measure, denoted \( \mathbb{B} \).

\subsubsection{Forward Affinity:}
Given an attributed graph \( G \), a vertex \( v_i \), and a restart probability \( \alpha \) with \( 0 < \alpha < 1 \), a \textit{forward random walk} begins at \( v_i \).  
When the walker is currently located at vertex \( v_l \), it will either stop at that vertex with probability \( \alpha \) or, with probability \( 1 - \alpha \), traverse one outgoing edge chosen uniformly at random to an out-neighbour of \( v_l \) contained in \( E_V \).

Once the random walk terminates at a node \( v_l \), an edge in \( E_R \) is randomly selected to connect to an attribute \( r_j \) based on the probability \( \mathbf{R}_r[v_l, r_j] \), where \( \mathbf{R}_r[v_l, r_j] \) represents a normalized edge weight as defined in Equation~\ref{eq:nor}. Upon reaching an attribute, we check whether it satisfies predefined rules. If a rule is satisfied, we examine the rule’s left-hand side (LHS). If the LHS contains a single literal, we directly perform a \textbf{\textit{rule jump}} to the right-hand side (RHS). For cases where the LHS contains multiple literals, we first propagate within the LHS before jumping to the RHS. This process yields more \textit{node-to-attribute pairs} \( (v_i, r_j) \) than previous random walks methods \cite{pane2}, and these pairs are collected in the set \( S_f \).

Assume that every vertex \( v_i \) contributes \( n_r \) sampled node–attribute pairs. Under this setting, the cardinality of \( S_f \) becomes \( n_r \cdot n + \delta \), where \( n \) is the number of vertices in \( G \) and \( \delta \) records extra pairs generated by rule jumps.  
Let \( p'_f(v_i, r_j) \) denote the probability that a forward random walk starting from \( v_i \) (with rule jumps enabled) outputs the pair \( (v_i, r_j) \).  
We therefore define the \textit{forward affinity} between \( v_i \) and \( r_j \) as

\begin{equation} \label{eq:f_aff}
    \
    \mathbb{F}[v_i, r_j] = \log \left( \frac{n \cdot p'_f(v_i, r_j)}{\sum_{v_h \in V} p'_f(v_h, r_j)} + 1 \right)
    \
\end{equation}

To interpret this quantity, note that within \( S_f \) the marginal probabilities satisfy  
\( \mathbf{P}(v_i) = 1/n \),  
\( \mathbf{P}(r_j) = \dfrac{\sum_{v_h \in V} p'_f(v_h, r_j)}{n + \delta} \),  
and the joint probability is  
\( \mathbf{P}(v_i, r_j) = \dfrac{p'_f(v_i, r_j)}{n + \delta} \).  
Hence \( \mathbb{F}[v_i, r_j] \) can be regarded as a \textit{Shifted PMI} (SPMI) score, where the constant \( +1 \) guarantees non-negativity. A larger value indicates a stronger tendency for the node–attribute pair to co-occur in \( S_f \).

\subsubsection{Backward affinity:} 
For a directed, attributed graph \( G \), an attribute \( r_j \), and a restart probability \( \alpha\;(0<\alpha<1) \), a \textit{backward random walk} is defined as follows.  
First, a vertex \( v_l \) is selected according to the distribution \( \mathcal{R}_c[v_l,r_j] \) given in Equation~\ref{eq:nor}. The walk then proceeds from \( v_l \); at each step it halts at the current node with probability \( \alpha \) or traverses a uniformly sampled outgoing edge with probability \( 1-\alpha \).  
Whenever the walk stops at vertex \( v_i \), we record the pair \( (r_j, v_i) \) and insert it into the multiset \( \mathcal{S}_b \).  
Sampling \( n_r \) such pairs for every attribute yields \( |\mathcal{S}_b| = n_r \cdot d \), where \( d \) is the number of distinct attributes.

Let \( p_b(v_i, r_j) \) be the probability that a backward walk starting from \( r_j \) terminates at \( v_i \). Inside \( \mathcal{S}_b \) we have  
\( \mathbf{P}(r_j) = 1/d \),  
\( \mathbf{P}(v_i) = \sum_{r_h \in \mathcal{R}} p_b(v_i, r_h) / d \),  
and \( \mathbf{P}(v_i,r_j) = p_b(v_i, r_j) \).  
By the Shifted PMI (SPMI) scheme, the \textit{backward affinity}~\cite{pane2} is defined as

\begin{equation}\label{eq:b_aff}
    \mathbb{B}[v_i, r_j] = \log \left( \frac{d \cdot p_b(v_i, r_j)}{\sum_{r_h \in \mathcal{R}} p_b(v_i, r_h)} + 1 \right).
\end{equation}

\subsubsection{Objective Function:}
Given the rule-based forward affinities \( \mathbb{F}[v_i,r_j] \) and backward affinities \( \mathbb{B}[v_i,r_j] \) (Eqs.~\ref{eq:f_aff}--\ref{eq:b_aff}), we learn compact embeddings under a space budget \( k \).  
Each vertex \( v_i \) is assigned two vectors in \( \mathbb{R}^{k/2} \): a forward embedding \( \mathbf{X}_f[v_i] \) and a backward embedding \( \mathbf{X}_b[v_i] \).  
Every attribute \( r_j \) obtains a vector \( \mathbf{Y}[r_j] \in \mathbb{R}^{k/2} \).  
The parameters are obtained by minimising the squared reconstruction error:

\begin{align}
    \mathcal{O} = \min_{\mathbf{X}_f, \mathbf{Y}, \mathbf{X}_b} \sum_{v_i \in V} \sum_{r_j \in \mathcal{R}} & \left( \left( \mathbb{F}[v_i, r_j] - \mathbf{X}_f[v_i] \cdot \mathbf{Y}[r_j]^{\top} \right)^2 \right. \notag \\
    & \left. + \left( \mathbb{B}[v_i, r_j] - \mathbf{X}_b[v_i] \cdot \mathbf{Y}[r_j]^{\top} \right)^2 \right).
\end{align}

The first term aligns the forward embeddings with forward affinities, while the second term enforces an analogous alignment for the backward direction, ensuring that the learned representations faithfully encode bidirectional node–attribute relationships within the prescribed dimensionality bound.

\subsection{Item Recommendation Module}
The enriched node and attribute embeddings generated from rule-based random walks are subsequently integrated into a GCN model adapted from~\cite{lightgcn}. This model leverages both structural and semantic information from the embeddings to predict user-item interactions more accurately. Specifically, we input both the forward and backward embedding vectors for each node, allowing the GCN model to capture transitive 
dependencies and attribute-based similarities. This integration enhances the model's ability to recommend relevant items to users based on their embedding proximity within the property graph.

To prepare the forward and backward embeddings for input into the GCN, we perform a series of operations, combining weighted summation, normalization, and scaling in a unified transformation. 
The combined embedding \( \mathbf{X}_{\text{final}} \in \mathbb{R}^{n \times d} \) for each node is computed as follows:

\begin{equation}
    \mathbf{X}_{\text{final}} = \frac{\alpha \cdot \mathbf{X}_f \odot \mathbf{Y} + \beta \cdot \mathbf{X}_b \odot \mathbf{Y} - \mu_{\text{sum}}}{\sigma_{\text{sum}}} \cdot \sqrt{\frac{2}{d_{\text{in}} + d_{\text{out}}}}
\end{equation}

We feed the normalized and scaled embeddings \( \mathbf{X}_{\text{final}} \) into the GCN model as the input layer. For training, we employ a pairwise loss function based on BPR, aiming to maximize the ranking of observed user-item interactions over unobserved ones. The optimization objective is defined as:

\begin{equation}
    L_{BPR} = - \sum_{u=1}^{M} \sum_{i \in N_u} \sum_{j \notin N_u} \ln \sigma (\hat{y}_{ui} - \hat{y}_{uj}) + \lambda \| \mathbf{E}^{(0)} \|^2,
\end{equation}

where $\sigma$ denotes the sigmoid function, $\hat{y}_{ui}$ and $\hat{y}_{uj}$ denote the predicted scores for an observed item $i$ and an unobserved (negative) item $j$ with respect to user $u$, respectively. 
Here, $M$ is the number of users, $N_u \subseteq I$ is the set of items user $u$ has interacted with, and $j \notin N_u$ indicates that $j$ is randomly sampled from the set of items not interacted with by $u$. This negative sampling process generates implicit contrastive pairs to guide the model toward ranking positive interactions higher than unobserved ones. The term $\lambda \| \mathbf{E}^{(0)} \|^2$ denotes $L_2$ regularization applied to the GCN input embeddings $\mathbf{E}^{(0)}$, constructed by combining rule-guided forward and backward node representations.
We utilize the Adam optimizer for training, with mini-batch sampling to efficiently learn the model parameters.

This optimization approach, focusing on ranking-based recommendation, allows the GCN model to leverage both structural and attribute-based information from the embeddings to provide improved recommendations.

\section{Experiments}
\subsection{Experimental Setup}
\begin{table}[h]
\centering
\caption{Summary of dataset statistics.}
\label{tab:dataset_statistics}
\begin{tabular}{llrrrrr}
\toprule
\textbf{Name} & \textbf{Type} & \(\mathbf{|V|}\) & \(\mathbf{|E_V|}\) & \(\mathbf{|R|}\) & \(\mathbf{|E_R|}\) & \textbf{Sparsity} \\
\midrule
Facebook & undirected & 4,039 & 88,234 & 1,283 & 33,301 & 0.00544 \\
Blogcatalog & undirected & 5,196 & 343,486 & 8,189 & 369,435 & 0.01272 \\
Flickr & undirected & 7,575 & 479,476 & 12,047 & 182,517 & 0.00836 \\
Citeseer & directed & 3,312 & 4,660 & 3,703 & 105,165 & 0.00053 \\
\bottomrule
\end{tabular}
\end{table}

Table \ref{tab:dataset_statistics} summarizes the datasets used in our experiments. The Facebook dataset consists of ego-networks where users interact with friends, and attributes are extracted from user profiles. The BlogCatalog dataset is a social network of bloggers with keywords as attributes. The Flickr dataset includes photos, with tags as attributes. Citeseer, a directed citation network, is used as a benchmark. $|V|$ and $|E_V|$ represent the number of nodes and edges, while $|R|$ and $|E_R|$ denote the number of attributes and node-attribute associations. The datasets vary in size and sparsity. Here, sparsity is defined as the ratio of the number of observed interactions to the total possible interactions between users and items. 
A lower sparsity value indicates a more incomplete interaction matrix, which often poses challenges for recommendations.

For our experiments, we split each dataset into 80\% for training and 20\% for testing. We compare our method against three baseline recommendation approaches: collaborative filtering (BPR-MF~\cite{bprmf}), GCN-based methods (LightGCN~\cite{lightgcn} and IMP-GCN~\cite{impgcn}), and attribute-enhanced methods (AF-GCN~\cite{afgcn}, \( \text{A}^{2}\text{-GCN}~\cite{a2gcn} \)). All methods are optimized using a pair-wise learning strategy, with Recall@20~\cite{recall} and NDCG@20~\cite{ndcg} as evaluation metrics. Recall@20 measures the proportion of relevant items in the top 20 recommendations out of all relevant items a user interacted with. NDCG@20, denoting the normalized discounted cumulative gain, evaluates the ranking quality by rewarding relevant items positioned higher in the list, with scores discounted logarithmically for lower ranks, normalized against the best possible ranking. 

In implementation, we used TensorFlow, the Adam optimizer, a mini-batch size of 2048 (1024 for Citeseer), and explored learning rates from {0.01, 0.001, 0.0001}. The L2 regularization coefficient was searched within $[10^{-5}$, $10^{-2}]$. Latent vector dimensions were set to 64, stopping and validation strategies followed those used in LightGCN~\cite{lightgcn}.
 
\subsection{Performance Comparison}

\begin{table}[t]
\centering
\caption{Performance of our RAE model and the competitors over four datasets. Values are reported as percentages (‘\%’ omitted).}
\label{tab:performance_comparison}
\begin{tabular}{llcccccccc}
\toprule
\textbf{Type} & \textbf{Method} 
  & \multicolumn{2}{c}{\textbf{Facebook}} 
  & \multicolumn{2}{c}{\textbf{Blogcatalog}} 
  & \multicolumn{2}{c}{\textbf{Flickr}} 
  & \multicolumn{2}{c}{\textbf{Citeseer}} \\
\cmidrule(lr){3-4} \cmidrule(lr){5-6} \cmidrule(lr){7-8} \cmidrule(lr){9-10}
 &  & \textbf{R@20} & \textbf{N@20} 
      & \textbf{R@20} & \textbf{N@20} 
      & \textbf{R@20} & \textbf{N@20} 
      & \textbf{R@20} & \textbf{N@20} \\
\midrule
\multirow{1}{*}{\textbf{CF}} 
  & BPR-MF      & 0.2342 & 0.2706 & 0.0976 & 0.1257 & 0.1372 & 0.1428 & 0.0609 & 0.0474 \\

\midrule
\multirow{2}{*}{\textbf{GCN}} 
  & LightGCN    & 0.2462 & 0.3132 & 0.1406 & 0.1249 & 0.2092 & 0.1696 & \underline{0.1246} & 0.0839 \\
  & IMP-GCN     & 0.2411 & 0.1578 & 0.1493 & 0.1361 & 0.1537 & 0.1537 & 0.1152 & \underline{0.0879} \\

\midrule
\multirow{2}{*}{\textbf{Attr.}} 
  & AF-GCN      & 0.2516 & 0.3173 & \underline{0.1640} & \underline{0.1495} & \underline{0.2227} & \underline{0.1792} & 0.1127 & 0.0793 \\
  & \(A^2\text{-GCN}\) & \underline{0.2554} & \underline{0.3214} & 0.1429 & 0.1412 & 0.1987 & 0.1624 & 0.1169 & 0.0803 \\

\midrule
\multirow{2}{*}{\textbf{Our}}  
  & RAE         & \textbf{0.2695} & \textbf{0.3381} & \textbf{0.1717} & \textbf{0.1588} & \textbf{0.2346} & \textbf{0.1853} & \textbf{0.1615} & \textbf{0.1005} \\
  & Improvement & 5.52\% & 5.22\% & 4.72\% & 6.21\% & 5.36\% & 3.35\% & 29.64\% & 14.41\% \\

\bottomrule
\end{tabular}
\parbox{\textwidth}{\scriptsize \textbf{Note:} \textbf{Bold} and \underline{underline} indicate best and second-best results, respectively.}
\end{table}

We evaluate our RAE method against five baselines, with results summarized in Table \ref{tab:performance_comparison}. Three key findings emerge from this evaluation.

GCN-based approaches (LightGCN, IMP-GCN) consistently outperform BPR-MF across all datasets, highlighting the importance of graph-based modeling in capturing complex user-item relationships through high-order connectivity. This demonstrates the superiority of GCN over traditional methods, underscoring its potential for improving recommendation systems. Models incorporating attribute data (AF-GCN, $A^2$-GCN) show significant performance improvements over non-attribute baselines. However, their performance is limited by simplistic attribute integration, as seen with AF-GCN's reliance on \texttt{<users, items, attributes>} triples, which fail to fully leverage the rich semantic structures inherent in property graphs.

RAE achieves state-of-the-art performance, with substantial improvements over the best baseline (AF-GCN). Its rule-mining mechanism enriches embeddings through semantic pattern discovery, significantly enhancing performance in challenging datasets. These results confirm RAE's effectiveness as a robust solution for real-world recommendation systems.

\begin{figure}[t]
  \centering
  \includegraphics[width=\textwidth]{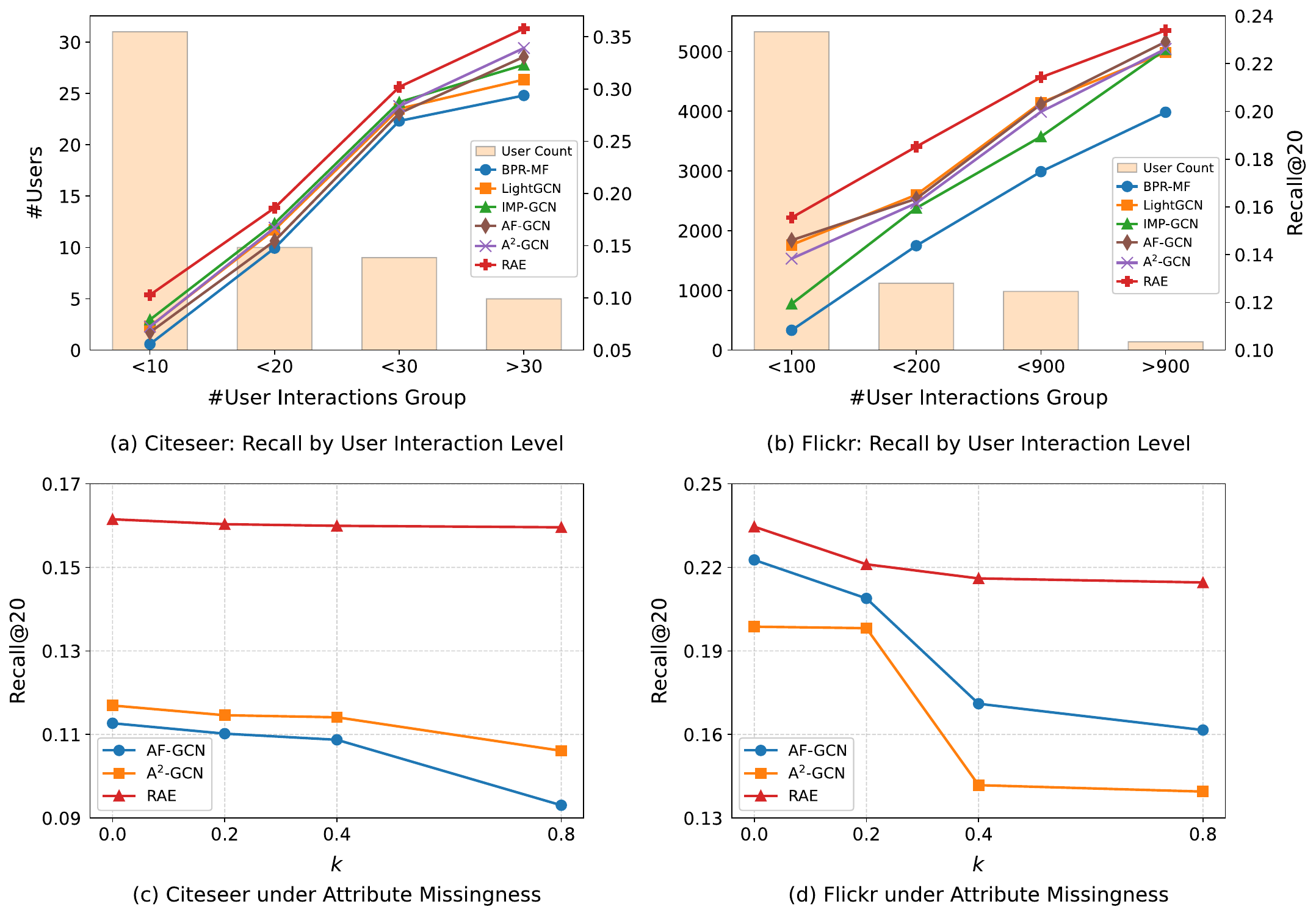}
  \caption{(a)-(b) Perfomance comparison of the sparsity distribution of the user on different datasets. (c)-(d) Impact of attributes missing with attribute removed randomly in ratio of {0, 0.2, 0.4, 0.8}. k is the ratio of removed attributes in each dataset.}
  \label{fig:expall}
\end{figure}

\subsection{Effects of Data Sparsity}

Attribute information plays a crucial role in addressing data sparsity issues in recommendation systems. To examine the effectiveness of RAE under varying levels of user interaction sparsity, we divided users into four groups based on their interaction counts: fewer than 100, 200, 900, and more than 900 for the Flickr dataset, and fewer than 10, 20, 30, and more than 30 for the Citeseer dataset. Fig. \ref{fig:expall} (a)-(b) presents the Recall@20 results for each user group on Citeseer and Flickr.

In both datasets, models that leverage attribute information, such as AF-GCN and $A^2$-GCN, consistently outperform collaborative filtering methods like BPR-MF, highlighting the benefit of attributes in sparse settings. GCN-based models, such as LightGCN and IMP-GCN, also show enhanced performance compared to traditional CF methods, indicating the effectiveness of utilizing high-order connectivity within graph structures.

RAE consistently surpasses all baseline models across all sparsity levels in both datasets. This indicates that the rule-driven attribute embedding in RAE is particularly effective for capturing relationships in sparse data. For instance, on the Citeseer dataset, RAE achieves notable improvements in each interaction group, demonstrating its resilience and effectiveness even when interactions are extremely limited. Interestingly, the performance of RAE improves as the number of interactions increases. For example, in Citeseer, RAE’s relative improvement over the best baseline method grows incrementally across groups with higher interaction counts. This trend suggests that while RAE is effective in very sparse scenarios, the presence of additional interactions enhances its ability to propagate attribute information within the graph structure, leading to even greater performance gains.

Overall, these results demonstrate that RAE effectively mitigates the challenges posed by data sparsity, outperforming baseline methods consistently by leveraging both attribute information and the graph structure. This robust performance in sparse datasets highlights RAE’s potential for enhancing recommendation accuracy in real-world applications where user interactions are often limited.

\subsection{Effects of the Attribute Missing Problem}
In this section, we investigate the effects of missing attributes on attribute-aware recommendation methodologies. To simulate this scenario, we randomly removed attribute labels from the datasets at varying rates: 0\%, 20\%, 40\%, and 80\%. Specifically, a 20\% missing rate indicates that 20\% of the attribute labels were randomly omitted from the dataset. Due to constraints on space, we focus our presentation of results on the Citeseer and Flickr datasets, as shown in Fig. \ref{fig:expall} (c)-(d). Notably, similar trends were observed across other datasets, reinforcing the robustness of our findings.

The results presented in Fig. \ref{fig:expall} (c)-(d) for the Citeseer and Flickr datasets, show that RAE consistently outperforms the baselines AF-GCN and $A^2$-GCN across all missing attribute levels. While performance declines as more attributes are missing, RAE experiences a notably smaller decline. For example, on Citeseer, RAE's Recall@20 drops slightly from 0.161 to 0.160 at 80\%, whereas AF-GCN and $A^2$-GCN show more significant reductions. Similar trends are observed on the Flickr dataset.

RAE's rule-driven embedding approach contributes to its robustness, as it effectively integrates attribute information and reduces dependency on complete data. This makes RAE particularly suitable for real-world recommendation tasks where attribute data may be incomplete, highlighting its practicality in such scenarios.

\subsection{Ablation Study}

\begin{table}[t]
\centering
\caption{Performance Comparison of Different RAE Variants on Various Datasets}
\label{tab:ablation}
\begin{tabular}{lcccccc}
\toprule
\textbf{Datasets} & \multicolumn{2}{c}{\textbf{Blogcatalog}} & \multicolumn{2}{c}{\textbf{Flickr}} & \multicolumn{2}{c}{\textbf{Citeseer}} \\
\cmidrule(lr){2-3} \cmidrule(lr){4-5} \cmidrule(lr){6-7}
\textbf{Methods} & \textbf{R@20} & \textbf{N@20} & \textbf{R@20} & \textbf{N@20} & \textbf{R@20} & \textbf{N@20} \\
\midrule
GCN$_{b}$ & 0.1406 & 0.1249 & 0.2092 & 0.1596 & 0.1246 & 0.0839 \\
RAE$_{h}$ & 0.1663 & 0.1537 & 0.2335 & 0.1789 & 0.1593 & 0.0991 \\
RAE$_{n}$ & 0.1659 & 0.1525 & 0.2322 & 0.1827 & 0.1575 & 0.0961 \\
RAE$_{u}$ & 0.1642 & 0.1485 & 0.2234 & 0.1792 & 0.1557 & 0.0636 \\
RAE & \textbf{0.1717} & \textbf{0.1588} & \textbf{0.2346} & \textbf{0.1853} & \textbf{0.1615} & \textbf{0.1005} \\
\bottomrule
\end{tabular}
\parbox{\textwidth}{\scriptsize Note: \textbf{Bold} indicates the best performance in each column.}
\end{table}

We analyze the contribution of different components to the performance of our RAE model by comparing it with the following variants: \textbf{GCN$_{b}$}, which applies only GCN to the graph structure without rule-based random walks or attribute embeddings; \textbf{RAE$_{h}$}, which uses half of the available rules for the rule-based random walk to assess the impact of partial rule application; \textbf{RAE$_{n}$}, which employs regular random walks without rule guidance, serving as an embedding baseline; and \textbf{RAE$_{u}$}, where the random walk embeddings are fed directly into the GCN without further refinement.

The results, presented in Table \ref{tab:ablation}, reveal that \textbf{RAE$_{h}$} outperforms both \textbf{GCN$_{b}$} and \textbf{RAE$_{u}$}, highlighting the significant enhancement brought by rule-based random walks, even with partial rule information. The improvement of RAE over \textbf{RAE$_{u}$} confirms the importance of embedding refinement after the random walk. \textbf{RAE$_{n}$}, which uses traditional random walks, performs better than \textbf{GCN$_{b}$} but still falls short of \textbf{RAE$_{h}$} and the complete RAE model, emphasizing the added value of rule-based guidance.

In summary, the ablation study confirms that each component of RAE, including rule-based random walks and embedding refinement, contributes significantly to its superior performance, validating our design choices.

\section{Conclusion}

In this work, we introduced \textbf{RAE}, a novel \textit{Rule-Driven Approach for Attribute Embedding}, designed to enhance recommendation systems by leveraging the semantic richness of property graphs. Through rule mining and rule-based random walks, RAE effectively utilizes attribute information within property graphs, overcoming the limitations of traditional Graph Convolutional Network (GCN) models that typically overlook the complex dependencies between attributes. Our extensive experiments demonstrate RAE’s superior performance in recommendation tasks across various datasets, showing substantial improvements in recall and robustness compared to state-of-the-art baselines.

Beyond improving accuracy, RAE’s design allows for better interpretability by tracing recommendations back to mined semantic rules, offering clearer insights into why items are suggested. This makes it particularly valuable in user-facing applications where transparency is critical. Overall, RAE provides a robust foundation for property graph-based attribute embedding and opens new directions for building more trustworthy, interpretable, and effective recommendation systems in real-world settings. Future work includes optimizing the computational efficiency of rule-based embedding and extending the framework to support dynamic attribute evolution in temporal or continuously evolving graphs.

\begin{credits}
\subsubsection{\ackname} The first author, affiliated with Huazhong Agricultural University, was funded by the Provincial Innovation and Entrepreneurship Training Program for Undergraduates under Grant No. S202310504323.  
The authors affiliated with institutions 1, 4, 5, and 6 were supported in part by the Hubei Key Research and Development Program of China under Grants 2024BBB055, 2024BAA008, and in part by the Fundamental Research Funds for the Chinese Central Universities under Grant 2662025XXPY005, and in part by the open funds of Hubei Three Gorges Laboratory under Grant SK232011.
\end{credits}
%
%
%
%

\end{document}